\documentclass[a4paper,11pt, eqno]{amsart}

\usepackage{mathrsfs, amssymb, eucal, amsmath, amsthm, amscd}
\usepackage{a4wide,latexsym,amssymb,amsthm,amsmath,amsfonts,url}
\usepackage{lipsum}
\usepackage{mathdesign}
\usepackage{cancel}
\usepackage{tikz}
\usepackage{tikz-3dplot}
\usepackage{pgfplots}
\usetikzlibrary{shapes}
\tdplotsetmaincoords{60}{110}

\usepackage{graphicx}
\usepackage{epstopdf}
\usepackage[centertags]{amsmath}
\usepackage{amsfonts}
\usepackage{newlfont}
\usepackage{amsfonts}
\usepackage{amssymb}
\usepackage{amsbsy}
\usepackage{amscd}
\usepackage{epsfig}
\usepackage{hyperref}
\usepackage{subfigure}

\def\frk{\frak}               % font for "Fraktur"

\def\Phi{{\frk n}}
\def\Phi{{\frk N}}

\def\opn#1#2{\def#1{\operatorname{#2}}} % to make operators
\opn\chara{char} \opn\length{\ell} \opn\pd{pd} \opn\rk{rk}
\opn\projdim{proj\,dim} \opn\injdim{inj\,dim} \opn\rank{rank}
\opn\depth{depth} \opn\grade{grade} \opn\height{height}\opn\coheight{coheight}
\opn\embdim{emb\,dim} \opn\codim{codim}

\opn\Tr{Tr} \opn\bigrank{big\,rank}
\opn\superheight{superheight}\opn\lcm{lcm}
\opn\trdeg{tr\,deg}%\emph{
\opn\reg{reg} \opn\lreg{lreg} \opn\ini{in} \opn\lpd{lpd}
\opn\size{size}\opn\bigsize{bigsize}
\opn\cosize{cosize}\opn\bigcosize{bigcosize}
\opn\sdepth{sdepth}\opn\sreg{sreg}
\opn\link{link}\opn\fdepth{fdepth}\opn\type{type}
\opn\GL{GL}
\opn\div{div} \opn\Div{Div} \opn\cl{cl} \opn\Cl{Cl}
\opn\Spec{Spec} \opn\Supp{Supp} \opn\supp{supp} \opn\Sing{Sing}
\opn\Ass{Ass} \opn\Min{Min}\opn\Mon{Mon} \opn\dstab{dstab} \opn\astab{astab}
\opn\Syz{Syz}
\opn\Ann{Ann} \opn\Rad{Rad} \opn\Soc{Soc}
\opn\Im{Im} \opn\Ker{Ker} \opn\Coker{Coker} \opn\Am{Am}
\opn\Hom{Hom} \opn\Tor{Tor} \opn\Ext{Ext} \opn\End{End}
\opn\Aut{Aut} \opn\id{id}

\opn\nat{nat}
\opn\pff{pf}%   \pf exists already
\opn\Pf{Pf} \opn\GL{GL} \opn\SL{SL} \opn\mod{mod} \opn\ord{ord}
\opn\Gin{Gin} \opn\Hilb{Hilb}\opn\sort{sort}
\opn\Proj{Proj}
\opn\aff{aff} \opn\con{conv} \opn\relint{relint} \opn\st{st}
\opn\lk{lk} \opn\cn{cn} \opn\core{core} \opn\vol{vol}
\opn\link{link} \opn\star{star}\opn\lex{lex}
\opn\gr{gr}

\opn\dirlim{\underrightarrow{\lim}}
\opn\inivlim{\underleftarrow{\lim}}
\def\Implies{\ifmmode\Longrightarrow \else
        \unskip${}\Longrightarrow{}$\ignorespaces\fi}
\def\implies{\ifmmode\Rightarrow \else
        \unskip${}\Rightarrow{}$\ignorespaces\fi}
\def\iff{\ifmmode\Longleftrightarrow \else
        \unskip${}\Longleftrightarrow{}$\ignorespaces\fi}

\let\:=\colon
\begin{document}

\title{RECOVERING THE COSMOLOGICAL CONSTANT FROM AFFINE GEOMETRY}
\author{Wladimir-Georges Boskoff; Salvatore Capozziello}
\address{Department of Mathematics \newline
\indent Ovidius University of Constanta, 900527, Constanta, Romania}
\email{boskoff@univ-ovidius.ro}
\address{Dipartimento di Fisica ``E. Pancini", \newline
 \indent Universit\`a degli Studi di Napoli
 \textquotedblleft{Federico II}\textquotedblright,\newline
\indent INFN Sez. di Napoli, Compl. Univ. di Monte S. Angelo, Edificio G, Via Cinthia, I-80126, Napoli, Italy,\newline
\indent Tomsk State Pedagogical University, ul. Kievskaya, 60, 634061 Tomsk, Russia.}
\email{capozziello@na.infn.it}

\begin{abstract}
A gravity theory without masses can be constructed in Minkowski spaces using a geometric Minkowski potential.
The related affine spacelike spheres can be seen as the regions of the Minkowski spacelike vectors characterized by a constant Minkowski gravitational potential. These spheres point out, for each dimension $n \geq 3$,  spacetime models, the de Sitter ones, which satisfy Einstein's field equations in  absence of matter. In other words, it is possible to generate geometrically  the cosmological constant. Even if  a lot of possible parameterizations have been proposed, each one highlighting some geometric and physical properties of the de Sitter space, we present here a new natural parameterization which reveals the intrinsic geometric nature of  cosmological constant relating it with the invariant affine radius  coming from the so called  Minkowski-Tzitzeica surfaces theory.

\end{abstract}

\maketitle

AMSC: 83C40, 83C05, 83C10.
\medskip

\medskip

\section{INTRODUCTION}

$\medspace$

$\medspace$

\medskip
The debate on the nature of cosmological constant is crucial both from cosmology and fundamental physics points of view.  It is related to the thorny problem of vacuum state in quantum gravity, with early time inflation as well as the recently observed accelerated expansion of the Hubble flow. In all these cases, the requested energy range is very different and then we need a vacuum state that is also  evolving, in a sense that any epoch requires  a suitable definition of the cosmological constant. This constitute the so called {\it Cosmological Constant Problem} \cite{Weinberg}. Several solutions can be invoked to solve it (see e.g.\cite{luongo,martin}) however no final answer is available at the moment. 

Besides this genuinely "physical approach" to the problem, geometric formulations of it have been proposed in order to answer  the old question  if cosmological constant is "geometry" (and then it has to be considered in the l.h.s. of Einstein's equations) or it is  "energy-matter" (and it has to be considered  into the r.h.s. of the field equations as a source). 
The two points of view, also if apparently equivalent,  have   enormous consequences and they  have  feedback in observational astrophysics and cosmology, see, for example,  \cite{Capozziello,Salvatore1,Ester,dunsby, alan,Nojiri:2006, Capozziello:2006,Report,vasilis}. 

Despite this situation, various approaches have been proposed to the parameterization of de Sitter spacetime \cite{Coxeter,Hartman,Hawking,Houches}.  

Here we propose a picture  based on the affine properties of Minkowski spacetime which allows  us to naturally deal with  the cosmological constant from a geometric point of view.
In a Minkowski $n$-dimensional space,  a gravitational force naturally appears. It is  determined by the origin $O$ which acts on each point $P$ if the vector $\stackrel{\rightarrow}{OP}$ is a spacelike one. This gravitational force $F$ is not induced by masses, but only by the geometric structure of the Minkowski space. The theory follows by assuming a gravitational Minkowski field $A$ and a Minkowski gravitational potential $\varPhi$. Theorems like  in  classical mechanics, such as  $\bigtriangledown_M \varPhi=-A$ or the Minkowski field equation in the form $\bigtriangledown_M^2 \varPhi=0$, can be immediately derived.\\

In this framework, for  each dimension,  we expect to find hypersurfaces whose metrics satisfy the  Einstein field equations (see \cite{Burke,Callahan,Moore1}) $$R_{ij}-\dfrac{1}{2}Rg_{ij}+\Lambda g_{ij}=\dfrac{8\pi G}{c^4}T_{ij} \eqno(1.1)$$ in the absence of standard matter-energy, that is $T_{ij}=0$.\\ 

The de Sitter spacetime is a remarkable example in this perspective. In the previous Minkowski geometric gravitational field,  the $(n-1)$-de Sitter spacetime is characterized by $\varPhi=-\dfrac{1}{a^{n-2}}$ 
because we are talking about the affine spacelike sphere $$X_0^2-X_1^2-...-X_{n-1}^2=-a^2\,,   \eqno(1.2)$$
where $a$ is a radius.
In the proposed parameterization,  one has $$R_{ij}+\dfrac{n-2}{a^2}g_{ij}=0,  \eqno(1.3)$$ 
that is $$R=-\dfrac{(n-1)(n-2)}{a^2}\,, \eqno(1.4)$$
for the Ricci curvature scalar. 
The Einstein field equations are satisfied for $$R=-\dfrac{(n-1)(n-2)}{a^2}, \;\; \Lambda=-\dfrac{(n-2)(n-3)}{2a^2}, \;\; T_{ij}=0.  \eqno(1.5)$$
However, even if the numerators are dimensional, the meaning of terms $$-\dfrac{(n-1)(n-2)}{a^2},\;\; -\dfrac{(n-2)(n-3)}{2a^2},\;\; \dfrac{(n-2)}{a^2}  \eqno(1.6)$$ is related to another geometric property available for this special hypersurface in the $n$-dimensional Minkowski space. In fact, we highlight a centro-affine invariant for the de Sitter spacetime suggested by the Tzitzeica affine differential theory, here in a Minkowski space.

The theory of Tzitzeica surfaces in Euclidean spaces was developed by the mathematician Gheorghe Tzitzeica,  whose adviser was Gaston Darboux at Sorbonne, Paris. Tzitzeica surfaces were the starting point of the affine differential geometry in which the differential invariants depend on metric. They are also preserved by the affine transformations (see \cite{Tzitzeica, doi}). \\
In the case of  Euclidean $3$-dimensional space, the definition of a Tzitzeica surface is  the following: one chooses a point $f(p)$ of the surface $f$ and  computes the Gaussian curvature $K_f(p)$ at that point. Then, we compute the distance, denoted by $d_f(p)$, from the origin $O(0,0,0)$ to the tangent plane to the surface at the point $f(p)$.
The surface is called a Tzitzeica surface, if the ratio $R^f(p):=\dfrac{K_f(p)}{d_f^4(p)}$ is a constant at each $p$, that is $R^f(p)=R^f$.  The constant $R^f$ is  an affine radius and becomes an intrinsic number attached to the surface.
The number is related to a metric relation but it is more than a metric relation: it  is a geometric invariant attached to the surface whose nature will be studied bellow. 
Suppose we  compute the affine radius $$R^f(p):=\dfrac{K_f(p)}{d_f^4(p)}  \eqno(1.7)$$ for the surface $f$ written locally in form $f(x,y)=(x,y, u(x,y))$ and $R^f(p)$ is a constant at each $p$.  \\
A centro-affine  transformation of the surface $f$ consists in the product between the surface $f$ and the $3 \times 3$ matrix $\mathbb{A}$, with $\det \mathbb{A} \neq 0$.
Therefore the surface obtained is now $$\bar{f}(x,y)=(a_{11}x+a_{21}y+a_{31} u(x,y), a_{12}x+a_{22}y+a_{32}u(x,y), a_{13}x+a_{23}y+a_{33}u(x,y),\eqno(1.8)$$ where the coefficients $a_{ij}$ are the components of the matrix $\mathbb{A}.$ For this surface $\bar{f}$, when we compute $R^{\bar{f}}(q)=\dfrac{K_{\bar{f}}(q)}{d_{\bar{f}}^4(q)}$,  we obtain a constant if the initial $R^f(p)=\dfrac{K_f(p)}{d_f^4(p)}$ is a constant. The constant is the same if and only if $\det \mathbb{A}=1$.

In fact the invariant is related to the constance of this ratio preserved by a centro-affine transformation. The definition depends on the normal $N$ to the surface, this one being involved both when we compute $K_f(p)=\dfrac{\det h_{ij}}{\det g_{ij}}$ and in the fourth power of the distance $d_f(p)$ from the origin to the tangent plane. \\
The previous observation allows us to extend the definition of the Tzitzeica surfaces to hypersurfaces in Euclidean $n$-dimensional spaces, the affine radius being the ratio $\dfrac{K_f(p)}{d_f^{(n+1)}(p)}$. \\
A comment is necessary at this point. Tzitzeica himself proved that the surface $xyz=1$ is a Tzitzeica surface and Calabi succeeded to generalize it to the hypersurface $x^1x^2...x^n=1$ \cite{Calabi}.
The main results  and some further conjectures can be found in \cite{Nomizu}.  
Some other results on Tzitzeica surfaces in Euclidean spaces are reported  in \cite{Agnew,Boskoff}.
A general theory about curves and surfaces in Minkowski spaces is presented in \cite{Lopez}, while the extension of the concept of Tzitzeica surfaces to Minkowski 3-dimensional spaces is presented in \cite{Bobe}.\\

In the present paper,  we discuss the Tzitzeica surfaces in $3$ and $4$-dimensional Minkowski spaces and we highlight the geometric nature of the affine Minkowski radius. The way we present the theory makes it easy to be generalized to Minkowski $n$-dimensional spaces.
Using an appropriate  parametrization,  it is possible to  prove that the $(n-1)$-de Sitter spacetime is a Minkowski-Tzitzeica hypersurface where the  cosmological constant is a consequence of the Minkowski affine radius.

The paper is organized as follows. In Sec. 2, we discuss how generate a gravitational force from the Minkowski geometry. This approach will reveal extremely important for the following considerations. Sec. 3 is devoted to  the definition of the Minkowski-Tzitzeica surfaces  which are the main ingredient  of the present approach because they establish the nature of the cosmological constant. The related affine radius is considered in Sec. 4 while the affine radius  in four dimensional Minkowski spaces is discussed in Sec. 5. The $(n-1)$-de Sitter spacetime, considered  as a Minkowski-Tzitzeica affine sphere, is presented in Sec.6. Discussion and conclusions are drawn in Sec.7.

\bigskip

\section{THE MINKOWSKI GEOMETRIC GRAVITATIONAL FORCE}

$\medspace$

$\medspace$

Let us denote  with $\mathbb{M}^n$ the Minkowski $n$-dimensional space, $n \geq 3$, endowed with the Minkowski product
$$ \left \langle a , b \right \rangle_M:= a_0 b_0-  \sum_{\alpha=1}^{n-1} a_{\alpha}b_{\alpha}   \eqno(2.1)$$
We choose to work with signature $ (+--...-)$. With respect to a given $b=(b_0, b_1, ..., b_{n-1})$, we consider all vectors $x=(x_0, x_1, ..., x_{n-1})$ such that $x-b$ is a spacelike vector, that is $\left \langle x-b , x-b \right \rangle_M < 0$.\\

We denote by $$r:=\sqrt{-(x_0-b_0)^2 + \sum_{\alpha=1}^{n-1} (x_{\alpha}-b_{\alpha}})^2    \eqno(2.2)$$ the Minkowski "length" of the spacelike vector $x-b$ and by $u=-\dfrac{1}{r}(x_0-b_0, x_1-b_1, ..., x_{n-1}-b_{n-1} )$ the unit vector of $b-x$.\\

Let us define the Minkowski geometric gravitational force by $$F^n_M:=\dfrac{1}{{n-1}} \ \dfrac{1}{r^{n-1}}\ u. \eqno(2.3)$$
If $$A^n_M:= \dfrac{n-2}{r^{n-1}}\ u   \eqno(2.4)$$ is, by definition, the geometric gravitational Minkowski field, we have the following "Minkowski-Newton second principle":
$$F^n_M=\dfrac{1}{(n-1)(n-2)}A^n_M.    \eqno(2.5)$$

Let us define the Minkowski gradient and the Minkowski Laplacian:
$$ \bigtriangledown_M :=\left(-\dfrac{\partial}{\partial x_0},   \dfrac{\partial}{\partial x_1}, ... , \dfrac{\partial}{\partial x_{n-1}}\right);   \eqno(2.6)$$
$$ \bigtriangledown_M^2 :=\left \langle \bigtriangledown_M , \bigtriangledown_M \right \rangle_M=\dfrac{\partial^2}{\partial x_0^2}- \dfrac{\partial^2}{\partial x_1^2}-...- \dfrac{\partial^2}{\partial x_{n-1}^2}.   \eqno(2.7)$$
For each dimension $n$,  we can define the Minkowski gravitational potential $$\varPhi_M^n :=-\dfrac{1}{r^{n-2}}.   \eqno(2.8)$$
The following computations 

$$\dfrac{\partial \varPhi_M^n}{\partial x_0}=(2-n)\dfrac{x_0-b_0}{r^n}; \ \ \dfrac{\partial \varPhi_M^n}{\partial x_{\alpha}}=(n-2)\dfrac{x_{\alpha}-b_{\alpha}}{r^n}, \ \alpha \in \{1,2,..., n-1\}; \eqno(2.9) $$

$$\dfrac{\partial^2 \varPhi_M^n}{\partial x_0^2}=(2-n)\dfrac{r^2+n(x_0-b_0)^2}{r^{n+2}}; \  \dfrac{\partial^2 \varPhi_M^n}{\partial x_{\alpha}^2}=(n-2)\dfrac{r^2-n(x_{\alpha}-b_{\alpha})^2}{r^{n+2}}, \ \alpha \in \{1,2,..., n-1\}. \eqno(2.10)$$

lead us to the following two theorems.\\

{\it Theorem 1: The Minkowski gradient of the Minkowski gravitational potential is the opposite of the Minkowski gravitational field. }

{\it Proof: } It is easy to check the equality $\bigtriangledown_M \varPhi_M^n=-A^n_M.$\\

{\it Theorem 2: The Minkowski Laplacian of the Minkowski gravitational potential is null.}\\

{\it Proof: } The same, it is easy to check  $\bigtriangledown_M^2 \varPhi_M^n=0.$\\

The last relation is the equation of the {\it Minkowski geometric gravitational field}.\\

In the case in which $b$ is the origin of the Minkowski space, the Minkowski unitary spacelike sphere can be thought as the set of $\mathbb{M}^n$ points described by the constant gravitational Minkowski potential $\varPhi_M^n=-1$. In this theory, at each dimension $n$, the Minkowski geometrical gravitational force and the Minkowski geometrical gravitational field have the dimension $\dfrac{1}{l^{n-1}}$. The Minkowski gravitational potential has the dimension $\dfrac{1}{l^{n-2}}$.  Here $l$ is a length.\\

We may conclude that, for  each dimension  in the Minkowski spacelike vector region, a natural geometric Minkowski gravity appears in  absence of matter. An equivalent of the Newton gravity theory can be constructed starting from the Minkowski geometric gravitational potential.  
The affine spacelike spheres can be seen as the regions of the Minkowski spacelike vectors characterized by a constant Minkowski gravitational potential. They highlight, at each dimension $n \geq 3$, a model of spacetime, the de Sitter one, which satisfies the Einstein field equations in  absence of matter, and it is now intuitive why it can happen.\\

\medskip

\section{THE MINKOWSKI-TZITZEICA SURFACES}

$\medspace$

$\medspace$

In  \cite{Bobe},  a way to extend the definition of a Tzitzeica surface to a Minkowski space is presented. Let us summarize  those results.  

In a Minkowski 3-dimensional space denoted by $\mathbb{M}^{3}$  we have\\
$\cdot $ the Minkowski product of vectors $$\left \langle a , b \right \rangle_M:= a_0 b_0- a_1b_1- a_2b_2.      \eqno(3.1)$$ 
$ \cdot$ the Minkowski crossproduct of vectors $$a \times_M b:= (a_1b_2-a_2b_1,a_0b_2-a_2b_0,a_1b_0-a_0b_1)    \eqno(3.2)$$ which can be easier understood from the formal determinant components,
$$
\left |
\begin{array}{cccc}
\stackrel{\rightarrow}{i} & -\stackrel{\rightarrow}{j} & -\stackrel{\rightarrow}{k} \\
a_0 & a_1 & a_2 \\
b_0 & b_1 & b_2 \\
\end{array}
\right |. \eqno(3.3)
$$

$\cdot$ A surface locally represented by $f:U=\overset{\circ}{U}\subset\mathbb{R}^{2}\longrightarrow\mathbb{M}^{3}$ having the form    
$f(x,y)=(x,y,u(x,y))$ with the metric $$ds^2=\left(1-\left(\dfrac{\partial u}{\partial x}\right)^2\right)dx^2-2\dfrac{\partial u}{\partial x}\dfrac{\partial u}{\partial y}dx dy-\left(1+\left(\dfrac{\partial u}{\partial y}\right)^2\right)dy^2   \eqno(3.4)$$
$\cdot$ The Gauss-Minkowski curvature $K^M_f\left( p\right)$ of $f$ at the point $f(p)$, where $p=\left( x,y\right) \in U,$ $$
K^M_f \left( p\right)= \dfrac{R_{1212}}{\det g_{ij}}=\dfrac{\dfrac{\partial^2 u}{\partial x^2}\dfrac{\partial^2 u}{\partial y^2} -\left(\dfrac{\partial^2 u}{\partial x \partial y}\right)^2 }{\left[\left( 
	\dfrac{\partial u}{\partial x}\right) ^{2}-\left( \dfrac{\partial u}{\partial y 
	}\right) ^{2}-1 \right]^2},
\eqno(3.5)$$ \\
$\cdot$ the equation of the tangent plane $\alpha $ at the point $f(p)$ is $$\alpha: (X-x)\left(-\dfrac{\partial u}{\partial x
}\right)-(Y-y)\left(\dfrac{\partial u}{\partial y
}\right)+(Z-u(x,y))=0    \eqno(3.6)$$
$\cdot$ the Minkowski distance, denoted $d^M_{f}(p)$, from the origin to the tangent plane of the surface $f$ at the point $f(p)$ computed after the formula
$$
d^M_{f}(p) =\dfrac{\left|x\dfrac{\partial u
	}{\partial x}+y\dfrac{\partial u}{\partial y} - u\left( x,y\right)\right|  }{\sqrt {\left|\left( 
		\dfrac{\partial u}{\partial x}\right) ^{2}-\left( \dfrac{\partial u}{\partial y 
		}\right) ^{2}-1 \right|}}.
\eqno(3.7)$$

An immediate consequence is $$ \dfrac{K^M_f \left( p\right)}{({d^M_{f}})^{4}\left( p\right)}
= \dfrac{\dfrac{\partial^2 u}{\partial x^2}\dfrac{\partial^2 u}{\partial y^2} -\left(\dfrac{\partial^2 u}{\partial x \partial y}\right)^2 }{\left(  x\dfrac{\partial u}{\partial x}+y%
	\dfrac{\partial u}{\partial y}-u\left( x,y\right)\right) ^{4}}. \eqno(3.8)$$ \\

\bigskip

Consider a matrix $\mathbb{A}\in \mathcal{M}_{3}(\mathbb{R}),$ such that $\det \mathbb{A}\neq 0$.

By definition, a centro-affine transformation of $f$ is a surface $\bar f:U=\overset{\circ }{U}\subset \mathbb{R}^{2}\longrightarrow \mathbb{M}^{3}$ given by the formula
$
\bar f(x,y)=f(x,y)\cdot \mathbb{A}.
$ where, explicitly 
$$
\bar f(x,y)=(x, y, u(x,y)) \cdot
\left (
\begin{array}{cccc}
a_{11} & a_{12} & a_{13} \\
a_{21} & a_{22} & a_{23} \\
a_{31} & a_{32} & a_{33} \\
\end{array}
\right )\eqno(3.9) 
$$
Therefore, $$\bar f(x,y)=\left(a_{11}x+a_{21}y+a_{31}u(x,y), a_{12}x+a_{22}y+a_{32}u(x,y), a_{13}x+a_{23}y+a_{33}u(x,y)   \right) .   \eqno(3.10)$$ 

Let us denote  by $\bar K^M_{\bar f}$ the Gauss-Minkowski curvature of 
$\bar f$ at the point $\bar{f}(p)$, where $p=\left( x,y\right)$,  and by $\Large {d^M}_{\bar f}$ the Minkowski distance from the origin to $\bar \alpha $, the tangent plane  of the surface $\bar f$ at the point $\bar f(p).$\\
One may observe that a centro-affine transformation changes the "shape" of a surface, changes the curvature of it at the new corresponding point and changes the distance between the origin and the tangent plane at the new corresponding point of the surface. Something remains invariant and the conclusion after the following theorem highlights that invariant. We can enunciate the following: \\
{\it Theorem (see \cite{Bobe}): $$\frac{\bar K^M_{\bar f}(p)}{\Large ({d^M}_{\bar f})^4(p)}=\dfrac{1}{(det \mathbb{A})^2}\cdot\frac{\dfrac{\partial^2 u}{\partial x^2}\dfrac{\partial^2 u}{\partial y^2} -\left(\dfrac{\partial^2 u}{\partial x \partial y}\right)^2}{\left( x\dfrac{\partial u}{\partial x}+y\dfrac{\partial u}{\partial y}-u(x,y)\right) ^{4}}=\dfrac{1}{(det \mathbb{A})^2}\cdot\dfrac{K^M_f \left( p\right)}{({d^M_{f}})^{4}\left( p\right)}
	.    \eqno(3.11)$$}

We prefer, instead of a very long computational proof of this theorem, to give some  geometric argument below. The theorem proved leads to a very important conclusion:\\

Starting from $f$, if  $\dfrac{K^M_f \left( p\right)}{({d^M_{f}})^{4}\left( p\right)}$ is a constant,  using the previous theorem,   we obtain that $\dfrac{\bar K^M_{\bar f}(p)}{\Large ({d^M}_{\bar f})^4(p)}$ is a constant.$\square$\\
With this result in mind, we can give the following \\

{\it Definition: In a Minkowski 3-dimensional space, a surface $f$ is called  a Minkowski-Tzitzeica surface if $\dfrac{K^M_f \left( p\right)}{({d^M_{f}})^{4}\left( p\right)}$ is a constant (see also \cite{Bobe})}\\

\medskip

\section{THE GEOMETRIC NATURE OF THE AFFINE RADIUS}

$\medspace$

$\medspace$

Let us consider  the ratio $\dfrac{K^M_f \left( p\right)}{{d^M}_{f}^{4}\left( p\right)}.$ Considering the formula (3.10), we can write the numerator as
$$ L:=\frac{\partial^2 u}{\partial x^2}\frac{\partial^2 u}{\partial y^2} -\left(\frac{\partial^2 u}{\partial x \partial y} \right)^2\eqno(4.1)$$ 
which can be written with respect to the vectors basis $$\dfrac{\partial f}{\partial x}=\left(1,0,\dfrac{\partial u}{\partial x}\right); \ \dfrac{\partial f}{\partial y}=\left(0,1,\dfrac{\partial u}{\partial y}\right)\eqno(4.2)$$ and four second order vectors $$ \dfrac{\partial^2 f}{\partial x^2}=\left(0,0,\dfrac{\partial^2 u}{\partial x^2}\right); \ \dfrac{\partial^2 f}{\partial y^2}=\left(0,0,\dfrac{\partial^2 u}{\partial y^2}\right); \ \dfrac{\partial^2 f}{\partial x \partial y}=\dfrac{\partial^2 f}{\partial y \partial x}=\left(0,0,\dfrac{\partial^2 u}{\partial x \partial y}\right)\eqno(4.3)$$ in the form

$$
L=
\left |
\begin{array}{cccc}
0 & 0 & \dfrac{\partial^2 u}{\partial x^2} \\
1 & 0 & \dfrac{\partial u}{\partial x } \\
0 & 1 & \dfrac{\partial u}{\partial y} \\
\end{array}
\right |\cdot 
\left |
\begin{array}{cccc}
0 & 0 & \dfrac{\partial^2 u}{\partial y^2} \\
1 & 0 & \dfrac{\partial u}{\partial x } \\
0 & 1 & \dfrac{\partial u}{\partial y} \\
\end{array}
\right | -
\left |
\begin{array}{cccc}
0 & 0 & \dfrac{\partial^2 u}{\partial x \partial y} \\
1 & 0 & \dfrac{\partial u}{\partial x } \\
0 & 1 & \dfrac{\partial u}{\partial y} \\
\end{array}
\right |\cdot
\left |
\begin{array}{cccc}
0 & 0 & \dfrac{\partial^2 u}{\partial y \partial x} \\
1 & 0 & \dfrac{\partial u}{\partial x } \\
0 & 1 & \dfrac{\partial u}{\partial y} \\
\end{array}
\right |.  \eqno(4.4)
$$
$L$ is in fact a difference of two products of $3$-volumes.\\ 

The denominator can be written, in the form 

$$
\left |
\begin{array}{cccc}
x & y & u(x,y) \\
1 & 0 & \dfrac{\partial u}{\partial x } \\
0 & 1 & \dfrac{\partial u}{\partial y} \\
\end{array}
\right |^4.  \eqno(4.5)
$$
Therefore is the $4^{th}$ power of a $3$-volume. The ratio $\dfrac{K^M_f \left( p\right)}{{d^M}_{f}^{4}\left( p\right)}$ has, at the denominator,  a dimension given as the second power of a $3$-volume. The cubic root $$\sqrt[3]{\dfrac{K^M_f \left( p\right)}{{d^M}_{f}^{4}\left( p\right)}} \eqno(4.6)$$  has as dimension $\dfrac{1}{l^2}$ and we will use this information later.\\
Furthermore, if we are looking at a centro-affine transformation, the vector bases  are 
$$\frac{\partial \bar f}{\partial x}=\left(a_{11}+a_{31}\frac{\partial u}{\partial x}, a_{12}+a_{32}\frac{\partial u}{\partial x} , a_{13}+a_{33}\frac{\partial u}{\partial x}  \right)\eqno(4.7)$$  
$$\frac{\partial \bar f}{\partial y}=\left(a_{21}+a_{31}\frac{\partial u}{\partial y}, a_{22}+a_{32}\frac{\partial u}{\partial y} , a_{23}+a_{33}\frac{\partial u}{\partial y}  \right)\eqno(4.8) $$ and the other   four  vectors are
$$\frac{\partial^2 \bar f}{\partial x^2}=\frac{\partial^2 u}{\partial x^2}\cdot(a_{31}, a_{32}, a_{33}), \ \frac{\partial^2 \bar f}{\partial x \partial y}=\frac{\partial^2 \bar f}{\partial y \partial x}=\frac{\partial^2 u}{\partial x \partial y}\cdot(a_{31}, a_{32}, a_{33}), \ \frac{\partial^2 \bar f}{\partial y^2}=\frac{\partial^2 u}{\partial y^2}\cdot(a_{31}, a_{32}, a_{33}).\eqno(4.9)$$
The  first determinant of the numerator is 
$$
\left |
\begin{array}{cccc}
a_{31}\dfrac{\partial^2 u}{\partial x^2} & a_{32}\dfrac{\partial^2 u}{\partial x^2} & a_{33}\dfrac{\partial^2 u}{\partial x^2} \\
a_{11}+a_{31}\dfrac{\partial u}{\partial x } & a_{12}+a_{32}\dfrac{\partial u}{\partial x } & a_{13}+a_{33}\dfrac{\partial u}{\partial x } \\
a_{21}+a_{31}\dfrac{\partial u}{\partial y } & a_{22}+a_{32}\dfrac{\partial u}{\partial y } & a_{23}+a_{33}\dfrac{\partial u}{\partial y} \\
\end{array}
\right |=det \mathbb{A} \cdot \dfrac{\partial^2 u}{\partial x^2},\eqno(4.10)
$$
therefore it becomes easy to see that the numerator has the value $det \mathbb{A}^2 \cdot \left(\dfrac{\partial^2 u}{\partial x^2}\dfrac{\partial^2 u}{\partial y^2} -\left(\dfrac{\partial^2 u}{\partial x \partial y} \right)^2\right). $\\
The denominator can be described by the following $4^{th}$ power of a determinant. The sign can be  neglected,
$$
\left |
\begin{array}{cccc}
a_{11}x+a_{21}y+a_{31}u(x,y) & a_{12}x+a_{22}y+a_{32}u(x,y) & a_{13}x+a_{23}y+a_{33}u(x,y) \\
a_{11}+a_{31}\dfrac{\partial u}{\partial x } & a_{12}+a_{32}\dfrac{\partial u}{\partial x } & a_{13}+a_{33}\dfrac{\partial u}{\partial x } \\
a_{21}+a_{31}\dfrac{\partial u}{\partial y } & a_{22}+a_{32}\dfrac{\partial u}{\partial y } & a_{23}+a_{33}\dfrac{\partial u}{\partial y} \\
\end{array}
\right |^4.\eqno(4.11)
$$
It is easy to compute the above determinant  using the standard  properties.\\
Finally, the denominator has the value $\det \mathbb{A}^4 \cdot \left(x\dfrac{\partial u}{\partial x}+y%
\dfrac{\partial u}{\partial y}-u\left( x,y\right)\right)^4. $ \\

We can conclude that the  Tzitzeica surfaces are related, both in the Minkowski case and in the initially considered Euclidean case, to the volume invariance.\\

A further  comment is useful at this point. Let us take into account  an affine transformation. The vectors $(1,0,0); \ (0,1,0), \ (0,0,1)$ transform into the vectors $(a_{11}, a_{12}, a_{13}); \ (a_{21}, a_{22}, a_{23}); \ (a_{31}, a_{32}, a_{33})$ described by the rows of the matrix $\mathbb{A}$. Therefore the initial volume, determined by the vectors $(1,0,0); \ (0,1,0), \ (0,0,1)$, that is $1$, is transformed into the volume $\det \mathbb{A}$. In the same way,  we can consider unit vectors on the initial axes, and then  the vectors $\dfrac{1}{\sqrt[3]{\det \mathbb{A}}}(a_{11}, a_{12}, a_{13});...$, which determine the unit volume after the centro-affine transformation. This means that the  affine radius is fully preserved, in  value, after a centro-affine transformation.\\
\bigskip

\section{THE AFFINE RADIUS  IN MINKOWSKI SPACES}

$\medspace$

$\medspace$

In a Minkowski $4-$dimensional space, denoted by $\mathbb{M}^4$, the $3$-surface we consider is $f(x,y,z)=(x,y,z,u(x,y,z))$,  with $(x,y,z)$ belonging to an open domain of $\mathbb{M}^3$.
The vectors that we take into account  are related to the tangent $3$-space, that is
$$\dfrac{\partial f}{\partial x}=\left(1,0,0, \dfrac{\partial u}{\partial x} \right); \  \dfrac{\partial f}{\partial y}=\left(0,1,0, \dfrac{\partial u}{\partial y} \right); \dfrac{\partial f}{\partial z}=\left(0,0,1, \dfrac{\partial u}{\partial z} \right), \eqno(5.1)$$
together with the other six second order derivatives 
$$ \dfrac{\partial^2 f}{\partial x^2}=\left(0,0,0, \dfrac{\partial^2 u}{\partial x^2} \right); \  \dfrac{\partial^2 f}{\partial x \partial y}=\left(0,0,0, \dfrac{\partial^2 u}{\partial x \partial y} \right); \ \dfrac{\partial^2 f}{\partial x \partial z}=\left(0,0,0, \dfrac{\partial^2 u}{\partial x \partial z} \right);\eqno(5.2)$$
$$ \dfrac{\partial^2 f}{\partial y^2}=\left(0,0,0, \dfrac{\partial^2 u}{\partial y^2} \right); \  \dfrac{\partial^2 f}{\partial y \partial z}=\left(0,0,0, \dfrac{\partial^2 u}{\partial y \partial z} \right); \ \dfrac{\partial^2 f}{\partial z^2 }=\left(0,0,0, \dfrac{\partial^2 u}{ \partial z^2} \right).\eqno(5.3)$$
If the surface is seen as a vector, there are involved  seven 4-determinants.  We have to discuss  the affine Minkowski radius. First of all, let us   compute $g_{ij}=\left\langle \dfrac{\partial f}{\partial x^{i}}\left(
x\right)
,\dfrac{\partial f}{\partial x^{j}}\left(
x\right) \right\rangle_M$. Therefore $$\det g_{ij}=-\left(\dfrac{\partial u}{\partial x}\right)^2 +\left(\dfrac{\partial u}{\partial y}\right)^2+\left(\dfrac{\partial u}{\partial z}\right)^2+1= \epsilon \left|\left(\dfrac{\partial u}{\partial x}\right)^2 -\left(\dfrac{\partial u}{\partial y}\right)^2-\left(\dfrac{\partial u}{\partial z}\right)^2-1  \right|,   \eqno(5.4)$$ where $\epsilon$ is the algebraic sign of $\det g_{ij}$.\\
Then, the $4$-Minkowski normal ${N}$ to the surface is related to the formal developing of the determinant 

$$
\left |
\begin{array}{ccccc}
\stackrel{\rightarrow}{i} & -\stackrel{\rightarrow}{j} & -\stackrel{\rightarrow}{k} & -\stackrel{\rightarrow}{l} \\
1 & 0 & 0 & \dfrac{\partial u}{\partial x } \\
0 & 1 & 0 & \dfrac{\partial u}{\partial y} \\
0 & 0 & 1 & \dfrac{\partial u}{\partial z} \\
\end{array}
\right |,\eqno(5.5)
$$
that is  

$${N}:=\dfrac{1}{\sqrt{\left|\left(\dfrac{\partial u}{\partial x}\right)^2 -\left(\dfrac{\partial u}{\partial y}\right)^2-\left(\dfrac{\partial u}{\partial z}\right)^2 -1\right|}}\left(\dfrac{\partial u}{\partial x}, -\dfrac{\partial u}{\partial y}, -\dfrac{\partial u}{\partial z}, 1\right).    \eqno(5.6)$$

If we denote $x_1=x, \ x_2=y, \ x_3=z$, the coefficients of the second fundamental form $h_{ij}$, in the case  the normal is a spacelike vector,  are  $$h_{ij}:=-\left\langle
{N},\dfrac{\partial ^{2}f}{\partial x^{i}\partial x^{j}} \right\rangle_M=\dfrac{\dfrac{\partial^2 u}{\partial x_i \partial x_j}}{\sqrt{\left|\left(\dfrac{\partial u}{\partial x}\right)^2 -\left(\dfrac{\partial u}{\partial y}\right)^2-\left(\dfrac{\partial u}{\partial z}\right)^2 -1\right|}
}  \eqno(5.7)$$ and the  3-Minkowski curvature at $f(p)$, where $p=(x,y,z)$, is

$$K^M_f(p):=-\dfrac{det h_{ij}}{det g_{ij}}=-\dfrac{\sum_{\sigma \in \Sigma_3 }\varepsilon(\sigma)\dfrac{\partial^2 u}{\partial x_1 \partial x_{\sigma(1)}}\dfrac{\partial^2 u}{\partial x_2 \partial x_{\sigma(2)}}\dfrac{\partial^2 u}{\partial x_3 \partial x_{\sigma(3)}}}{\left| \left(\dfrac{\partial u}{\partial x_1}\right)^2 -\left(\dfrac{\partial u}{\partial x_2}\right)^2-\left(\dfrac{\partial u}{\partial x_3}\right)^2 -1\right|^{\dfrac{5}{2}}}.  \eqno(5.8)$$
If the normal is a timelike vector, the sign $-$ in previous formulas becomes $+$. The distance from the origin to the tangent plane at $f(p)$ is 
$$d^M_f(p)=\dfrac{\left|\sum_{i=1}^{3}x_i\dfrac{\partial u}{\partial x_i}-u(x_1,x_2,x_3) \right|}{\sqrt{\left|\left(\dfrac{\partial u}{\partial x}\right)^2 -\left(\dfrac{\partial u}{\partial y}\right)^2-\left(\dfrac{\partial u}{\partial z}\right)^2 -1\right|}},   \eqno(5.9)$$
therefore the Minkowski affine radius is, modulo a sign, the ratio $$\dfrac{K^M_f(p)}{(d^M(p))^5}=-\dfrac{\sum_{\sigma \in \Sigma_3 }\varepsilon(\sigma)\dfrac{\partial^2 u}{\partial x_1 \partial x_{\sigma(1)}}\dfrac{\partial^2 u}{\partial x_2 \partial x_{\sigma(2)}}\dfrac{\partial^2 u}{\partial x_3 \partial x_{\sigma(3)}}}{\left|\sum_{i=1}^{3}x_i\dfrac{\partial u}{\partial x_i}-u(x_1,x_2,x_3) \right|^5}.  \eqno(5.10)$$
It is clear that $\dfrac{\partial^2 u}{\partial x_i \partial x_{\sigma(i)}}$ is the value of the determinant 

$$
\left |
\begin{array}{ccccc}
0 & 0 & 0 & \dfrac{\partial^2 u}{\partial x_i \partial x_{\sigma(i)}} \\
1 & 0 & 0 & \dfrac{\partial u}{\partial x } \\
0 & 1 & 0 & \dfrac{\partial u}{\partial y} \\
0 & 0 & 1 & \dfrac{\partial u}{\partial z} \\
\end{array}
\right |\eqno(5.11)
$$
and $\left|\sum_{i=1}^{3}x_i\dfrac{\partial u}{\partial x_i}-u(x_1,x_2,x_3) \right|$ is the absolute value of the determinant 

$$
\left |
\begin{array}{ccccc}
x & y & z & u(x,y,z) \\
1 & 0 & 0 & \dfrac{\partial u}{\partial x } \\
0 & 1 & 0 & \dfrac{\partial u}{\partial y} \\
0 & 0 & 1 & \dfrac{\partial u}{\partial z} \\
\end{array}
\right |.\eqno(5.12)
$$
At the numerator, we have six terms as products of $4$-determinants, therefore the dimension is the third power of a $4$-volume. At the denominator, we have the $5^{th}$ power of a $4$-volume, therefore the affine radius has, as dimension  at the denominator, the second power of a $4$-volume, therefore $$\sqrt[4]{\left|\dfrac{K_f \left( p\right)}{{d}_{f}^{5}\left( p\right)}\right|} \eqno(5.13)$$ is measured in $\dfrac{1}{l^2}$.\\

After a centro-affine transformation $\mathbb{A}\in \mathcal{M}_{4}(\mathbb{R})$ of the surface $f$, the connection between the two  affine radii is 

$$\frac{\bar K^M_{\bar f}}{\Large ({d^M}_{\bar f})^5}=\dfrac{(det \mathbb{A})^3}{(det \mathbb{A})^5}\cdot \frac{ K^M_{ f}}{\Large ({d^M}_{f})^5},\eqno(5.14)$$ therefore the same relation holds 
$$\frac{\bar K^M_{\bar f}}{\Large ({d^M}_{\bar f})^5}=\dfrac{1}{(det \mathbb{A})^2}\cdot \frac{ K^M_{ f}}{\Large ({d^M}_{f})^5}.\eqno(5.15)$$  According to these considerations,  it is clear that the result  can be extended to any   dimension.\\

\bigskip

\section{THE $(n-1)$-DE SITTER SPACETIMES AS   MINKOWSKI-TZITZEICA  AFFINE SPHERES}

$\medspace$

$\medspace$

\medskip

In the case $n=3$,  we choose to represent the $2$-surface $$X_0^2-X_1^2-X_2^2=-a^2   \eqno(6.1)$$ in the form 
$f:\mathbb{R}\times(-\pi,\pi)\longrightarrow \mathbb{M}^{3}$, $$f(t,x_1)=(a\sinh t, a\cosh t \cos x_1, a\cosh t \sin x_1).  \eqno(6.2)$$ 
Some computations lead to the metric $$ds_2^2=a^2dt^2-a^2\cosh^2 t \ dx_1^2.   \eqno(6.3)$$

The related non-zero Christoffel symbols are
$$ \Gamma^1_{01}=\Gamma^1_{10}=\tanh t, \  \Gamma^0_{11}=\cosh t \sinh t\eqno(6.4)$$
and then the Riemann tensor  component is 
$$
R_{101}^0=\frac{\partial \Gamma_{11}^0}{\partial t}-\frac{\partial \Gamma_{10}^0}{\partial x_1}+\Gamma _{s0}^0\Gamma_{11}^s-\Gamma_{s1}^0\Gamma_{10}^s=\cosh^2 t.\eqno(6.5)$$
It results $R_{0101}=g_{00}R_{101}^0=a^2\cosh^2 t$, that is $K^M_f=-\dfrac{1}{a^2}.$\\

The Minkowski normal is in fact
$${N}(t,x_1)=\dfrac{1}{a}f(t,x_1)\eqno(6.6)$$ \\
and this can be directly observed from the null Minkowski products 
$ \left\langle f, \dfrac{\partial f}{\partial t}\right\rangle_M;$  $\left\langle f , \dfrac{\partial f}{\partial x_1}\right\rangle_M$. 

According to the theory of surfaces in Minkowski spaces,  when the normal is a spacelike vector, the second fundamental form coefficients are\\ $$h_{ij}=\left\langle \dfrac{\partial N}{\partial x^{i}}
,\dfrac{\partial f}{\partial x^{j}} \right\rangle_M=\dfrac{1}{a}g_{ij}; \eqno(6.7)$$ and the Gauss-Minkowski curvature is computed with the formula $K_f^M:=-\dfrac{\det h_{ij}}{\det g_{ij}}=-\dfrac{1}{a^2}.$\\

Since we have ${N}(t,x_1)=\dfrac{1}{a}f(t, x_1)$, the Minkowski distance from the origin to the tangent plane has to be computed between the origin and the tangent point of the tangent plane at the surface. 
The result is the Minkowski radius $a$. \\

It is  $\dfrac{K^M_f}{\left({d_f^M}\right)^4}=-\dfrac{1}{a^6}$, that is the given Minkowski $2$-sphere is a Minkowski-Tzitzeica affine sphere.\\

For this $2$-de Sitter spacetime, according to Einstein theorem for surfaces $$R_{ij}=K^M_f \ g_{ij}   \eqno(6.8)$$ we have  $$R_{ij}+\dfrac{1}{a^2} \ g_{ij}=0. \eqno(6.9)$$ This last equation can be written as $$R_{ij}-\dfrac{1}{2}R \ g_{ij}=0,  \eqno(6.10)$$ that is $\Lambda=0$ and $T_{ij}=0$. $R$ is still depending on the centro-affine invariant $$\sqrt[3]{\left|\dfrac{K_f \left( p\right)}{{d}_{f}^{4}\left( p\right)}\right|}. \eqno(6.11)$$\\

In the case $n=4$, the $3$-de Sitter spacetime is the Minkowski spacelike sphere of $\mathbb{M}^4$ having the equation $$X_0^2-X_1^2-X_2^2-X_3^2=-a^2.    \eqno(6.12)$$
The usual parametrization is $$f(t,x_1,x_2)=(a\sinh t \cos x_2, a\cosh t \cos x_1 \cos x_2, a\cosh t \sin x_1 \cos x_2, a\sin x_2).  \eqno(6.13)$$
The metric is $$ds_3^2=\cos^2 x_2 (\ a^2 dt^2 -a^2\cosh^2 t  \ dx_1^2) -a^2dx_2^2,    \eqno(6.14)$$
 and it is easy to derive $$ds_3^2=\cos^2 x_2 \ ds_2^2 -a^2dx_2^2.    \eqno(6.15)$$
The non-zero Christoffel symbols are 

$$ \Gamma^0_{02}=\Gamma^0_{20}=-\tan x_2, \  \Gamma^0_{11}=\cosh t \sinh t,\eqno(6.16)$$

$$ \Gamma^1_{01}=\Gamma^1_{10}=\tanh t, \  \Gamma^1_{12}=\Gamma^1_{21}=- \tan x_2,\eqno(6.17)$$

$$\Gamma^2_{00}=-\sin x_2 \cos x_2, \  \Gamma^2_{11}=\cosh^2 t \cos x_2\sin x_2.\eqno(6.18)$$
Now, if we compute $$R_{ii}=R^s_{isi}=\dfrac{\partial \Gamma^s_{ii}}{\partial x^s}-\dfrac{\partial \Gamma^s_{is}}{\partial x^i}+\Gamma^h_{ii}\Gamma^s_{hi}-\Gamma^h_{is}\Gamma^s_{hi},\eqno(6.19)$$ we find $$R_{00}=-2\cos^2 x_2; \ R_{11}=2\cosh^2 t \cos^2 x_2; \ R_{22}=2.\eqno(6.20)$$ The other Ricci symbols are null, $ \ R_{ij}=0, i \neq j.$
Therefore $$R_{ij}+\dfrac{2}{a^2} \ g_{ij}=0.  \eqno(6.21)$$
If we compute $R:=R^i_i$, taking into account $R^i_j=g^{is}R_{sj}$,  it results $R=-\dfrac{6}{a^2}$.\\
The l.h.s. of Einstein's field equations becomes $$R_{ij}-\dfrac{1}{2}\left(-\dfrac{6}{a^2}\right)g_{ij}+\Lambda g_{ij}.  \eqno(6.22)$$
If we choose $\Lambda=-\dfrac{1}{a^2}$, it is  
$$R_{ij}+\dfrac{2}{a^2}\ g_{ij}=0 .\eqno(6.23)$$ 
The de Sitter spacetime presented above satisfies the Einstein field equations  (1.1) for $R=-\dfrac{6}{a^2}, \ \Lambda=-\dfrac{1}{a^2} $ and $T_{ij}=0$. A spacetime without matter appears as we expected.\\

Now again the  normal is $${N}=(\sinh t  \cos x_2, \cosh t \cos x_1 \cos x_2, \cosh t \sin x_1 \cos x_2,  \sin x_2 )=\dfrac{1}{a}f,\eqno(6.24)$$
therefore in this case, $$K^M_f(p):=-\dfrac{\det h_{ij}}{\det g_{ij}}:=-\dfrac{1}{a^3}.\eqno(6.25)$$ 
It is easy to see that $\left\langle
f ,\dfrac{\partial f}{\partial x^{i}} \right\rangle_M=0, \ i \in \{0,1,2\}$, that  is $d^M_f(p)=a$. The Minkowski sphere becomes a Minkowski-Tzizeica  affine sphere because $$\dfrac{K^M_f(p)}{(d^M_f(p))^5}=-\dfrac{1}{a^8}.\eqno(6.26)$$ 
According to both the geometric affine invariant meaning and the fact that the quantity is measured in $\dfrac{1}{l^2}$, the geometric quantity $$ \sqrt[4]{\left|\dfrac{K^M_f \left( p\right)}{({d}_{f}^{M}\left( p\right))^5}\right|} \eqno{(6.27)}$$ is related to  the dimensional constants of the de Sitter spacetime. \\

In the case $n=5$,  the parameterization is 
$$
\left \{
\begin{array}{l}
X_0=a\sinh t \cos x_2 \cos x_3\\
X_1=a\cosh t \cos x_1 \cos x_2 \cos x_3\\
X_2=a\cosh t \sin x_1 \cos x_2 \cos x_3\\
X_3=a\sin x_2 \cos x_3\\
X_4=a\sin x_3\\
\end{array}
\right.    \eqno(6.28)
$$
The metric generated  by this parametrization is 
$$ds_4^2=\cos^2 x_3 (a^2\cos^2x_2 dt^2-a^2\cosh^2 t\cos^2x_2 dx_1^2-a^2dx_2^2)-a^2dx_3^2.   \eqno(6.29)$$
In the same way, as in the previous case, we obtain 

$$ds_4^2=\cos^2 x_3 \ ds_3^2-a^2dx_3^2.   \eqno(6.30)$$
Again, if we compute $$R_{ii}=R^s_{isi}=\dfrac{\partial \Gamma^s_{ii}}{\partial x^s}-\dfrac{\partial \Gamma^s_{is}}{\partial x^i}+\Gamma^h_{ii}\Gamma^s_{hi}-\Gamma^h_{is}\Gamma^s_{hi},\eqno(6.31)$$ we find $$R_{00}=-3\cos^2x_2 \cos^2 x_3, \ R_{11}=3\cosh^2 t \cos^2x_2 \cos^2 x_3, \ R_{22}=3 \cos^2 x_3, \ R_{33}=3, \eqno(6.32)$$ that is

$$R_{ij}+\dfrac{3}{a^2}\ g_{ij}=0,   \eqno(6.33)$$ 
which leads to 
$$R=-\dfrac{12}{a^2},  \  \Lambda=-\dfrac{3}{a^2},  \ T_{ij}=0  \eqno(6.34)$$
for Einstein's field equations in this case.

As in the previous $n=4$ case, the Minkowski normal to the 4-hypersurface is $${N}(t,x_1,x_2,x_3)=\dfrac{1}{a}f(t,x_1,x_2,x_3),\eqno(6.35)$$ a spacelike vector.\\

It means that the Minkowski distance from the origin to the tangent 4-hyperplane at given point of the 4-hypersurface is $a$ and all the coefficients of the second fundamental form have the property $$h_{ij}=\dfrac{1}{a}g_{ij}.    \eqno(6.36)$$
Therefore $$K^M_f=-\dfrac{\det h_{ij}}{\det g_{ij}}=-\dfrac{1}{a^4}  \eqno(6.37)$$
and $$\dfrac{K^M_f(p)}{(d^M_f(p))^6}=-\dfrac{1}{a^{10}},  \eqno(6.38)$$
that is the absolute value of its $5^{th}$ root establish the invariant involved in the coefficients of Einstein's field equations.\\

In the general case the $(n-1)$-de Sitter spacetime is the Minkowski $(n-1)$-sphere determined by the ends of all the  spacelike vectors with Minkowski length $a$.\\ This is a hipersurface of the Minkowski $n$-dimensional space $\mathbb{M}^n$ having the algebraic equation $$X_0^2-X_1^2-...-X_{n-1}^2=-a^2.  \eqno(6.39)$$ 
The parameterization is
$$
\left \{
\begin{array}{l}
X_0=a\sinh t \cos x_2 \cos x_3 ....\cos x_{n-2} \\
X_1=a\cosh t \cos x_1 \cos x_2 \cos x_3 ....\cos x_{n-2}\\
X_2=a\cosh t \sin x_1 \cos x_2 \cos x_3 ....\cos x_{n-2} \\
X_3=a\sin x_2 \cos x_3 \cos x_4....\cos x_{n-2} \\
X_4=a\sin x_3 \cos x_4 ....\cos x_{n-2}\\
............................................\\
X_{n-2}=a\sin x_{n-3} \cos x_{n-2}\\
X_{n-1}=a\sin x_{n-2}\\
\end{array}
\right.  \eqno(6.40)
$$
This parameterization is meaningful  for $n\geq 5$.\\
For $n\geq 6$ we can denote $X_{0,n}:=X_0; \ X_{1,n}:=X_1; \ X_{n-1,n}:=X_{n-1}$ and it makes sense to write 

$$
\left \{
\begin{array}{l}
X_{0,n}=X_{0,n-1}\cos x_{n-2} \\
X_{1,n}=X_{1,n-1}\cos x_{n-2}\\
.............................\\
X_{n-2,n}=X_{n-2,n-1} \cos x_{n-2}\\
X_{n-1,n}=a\sin x_{n-2}\\
\end{array}
\right.  \eqno(6.41)
$$
with 
$$X_{0,n-1}^2-X_{1,n-1}^2-...-X_{n-2,n-1}^2=-a^2.  \eqno(6.42)$$
A direct consequence is 
$$X_{0,n-1}dX_{0,n-1}-X_{1,n-1}dX_{1,n-1}-...-X_{n-2,n-1}dX_{n-2,n-1}=0.  \eqno(6.43)$$
Using 
$$
\left \{
\begin{array}{l}
dX_{0,n}=dX_{0,n-1}\cos x_{n-2}-X_{0,n-1}\sin x_{n-2}dx_{n-2} \\
..............................................................................\\
dX_{n-2,n}=dX_{n-2,n-1} \cos x_{n-2}-X_{n-2,n-1}\sin x_{n-2}dx_{n-2}\\
dX_{n-1,n}=a\cos x_{n-2}dx_{n-2}\\
\end{array}
\right.  \eqno(6.44)
$$
and denoting by $$ds^2_k=dX^2_{0,k+1}-dX^2_{1,k+1}-...-dX^2_{k,k+1}  \eqno(6.45)$$ we obtain 
$$ds^2_{n-1}=a^2\cos^2 x_{n-2} \ ds^2_{n-2}-a^2dx^2_{n-2}, \ n\geq 6,   \eqno(6.46)$$
a formula which is the generalization of the formulas obtained for the previous cases $\ n=4$ and $n=5$.\\
In conclusion, in all cases we proved,  the metric is a diagonal one.\\

Let us now develop some further considerations.\\
If $$f(t,x_1,x_2,..,x_{n-2})=\left(X_{0,n-1}\cos x_{n-2}, ...,X_{n-2,n-1}\cos x_{n-2}, a\sin x_{n-2}\right)\eqno(6.47)$$ the direct consequence of  the above considerations is 
$ \left\langle f, \dfrac{\partial f}{\partial t}\right\rangle_M=0.$ \\
Another computation leads to $ \left\langle f, \dfrac{\partial f}{\partial x_k}\right\rangle_M=0,$ while $ \left\langle \dfrac{\partial f}{\partial x_k}, \dfrac{\partial f}{\partial t}\right\rangle_M=0$, $ \left\langle \dfrac{\partial f}{\partial x_k}, \dfrac{\partial f}{\partial x_j}\right\rangle_M=0$ are the consequences of the diagonal form of the metric and highlight the orthogonal frame of the tangent space at each point.\\
Therefore the Minkowski normal to the hypersurface is $${N}(t,x_1,...,x_{n-2})=\dfrac{1}{a}f(t,x_1,...,x_{n-2}),\eqno(6.48)$$ that is the Minkowski distance from the origin to the tangent hyperplane at a given point of the hypersurface is $a$ and all the coefficients of the second fundamental form are computed with the formula established for the case $n=2$, that is $$h_{ij}=\left\langle \dfrac{\partial N}{\partial x_i},\dfrac{\partial f}{\partial x_j}  \right\rangle_M, \eqno(6.49)$$ and then  $$h_{ij}=\dfrac{1}{a}g_{ij}.\eqno(6.50)$$

Since $ \left\langle N, N\right\rangle_M=-1<0$ we have
$$R_{ijkl}=-\left(h_{ik}h_{jl}-h_{il}h_{jk}\right).\eqno(6.51)$$ It results
$$R_{ijkl}=-\dfrac{1}{a^2}\left(g_{ik}g_{jl}-g_{il}g_{jk}\right), \  i,j,k,l \in \{0,1,...,{n-2}\}.   \eqno(6.52)$$

Therefore each sectional curvature is $$K=-\dfrac{1}{a^2}\eqno(6.53)$$
and
 
$$R^m_{jkl}=g^{mi}R_{ijkl}=-\dfrac{1}{a^2}\left(\delta^m_k g_{jl}-\delta^m_l g_{jk}\right).\eqno(6.54)$$
Finally $$R_{jl}=\sum_{m=0}^{n-2}R^m_{jml}=-\dfrac{1}{a^2} \sum_{m=0}^{n-2}(g_{jl}-\delta^m_l g_{jm})=-\dfrac{n-2}{a^2}g_{jl}.    \eqno(6.55)$$
Since $$R_{ij}+\dfrac{1}{2}(n-1)(n-2)\dfrac{1}{a^2}\ g_{ij}-\dfrac{(n-2)(n-3)}{2}\dfrac{1}{a^2} \ g_{ij}=R_{ij}+\dfrac{n-2}{a^2} \ g_{ij}=0  \eqno(6.56)$$ it results that if we choose
$$ \Lambda=-\dfrac{(n-2)(n-3)}{2}\dfrac{1}{a^2},   \eqno(6.57)$$ the previous metric satisfies the Einstein field equations (1.1) in the absence of matter, that is for $T_{ij}=0.$
At the same time,  $$ K^M_f:=-\dfrac{\det h_{ij}}{\det g_{ij}}=-\dfrac{1}{a^{n-1}}; \ d^M_f:=a. \eqno{(6.58)} $$ 
The generalization is immediate. In fact, the constants can be  written with respect to the Minkowski-Tzitzeica affine invariant 

$$ \sqrt[n]{\left|\dfrac{K^M_f \left( p\right)}{({d}_{f}^{M}\left( p\right))^{n+1}}\right|}=\dfrac{1}{a^2},\eqno(6.59)$$
i.e the cosmological constant is
$$  \Lambda =-\dfrac{(n-2)(n-3)}{2} \sqrt[n]{\left|\dfrac{K^M_f \left( p\right)}{({d}_{f}^{M}\left( p\right))^{n+1}}\right|},  \eqno(6.60)$$
the curvature scalar is
$$  R =-(n-1)(n-2) \sqrt[n]{\left|\dfrac{K^M_f \left( p\right)}{({d}_{f}^{M}\left( p\right))^{n+1}}\right|},   \eqno(6.61)$$
and the initial relation entering the Einstein field equations can be written in the form
$$R_{ij}+(n-2) \sqrt[n]{\left|\dfrac{K^M_f \left( p\right)}{({d}_{f}^{M}\left( p\right))^{n+1}}\right|}g_{ij}=0.   \eqno(6.62)$$
The last three equations are the most general results of this study.

\bigskip

\medskip

\section{DISCUSSION AND CONCLUSIONS}

\medspace

\medspace

In 1920 in Leiden, the cosmological spacetime structure  was the subject of a joint work between de Sitter and Einstein. De Sitter presented to Einstein, the so called de Sitter spacetime, as a curiosity: the result was that the Einstein field equations can produce  a dynamical spacetime  as a solution without matter-energy as source.\\ The present paper offers a possible answer to this issue  related to the possibility to generate gravity from  geometry   without mass-energy distribution: in this case, the de Sitter spacetime is characterized   as an hypersurface expressed assuming a  "constant Minkowski potential". 
\\
The properties of the de Sitter spacetime are highlighted by the new parameterization we presented here, whose metric coefficients are given in a recursive way, that is 

$$ds^2_{n-1}=a^2\cos^2 x_{n-2} \ ds^2_{n-2}-a^2dx^2_{n-2}, \ n\geq 4  \eqno(7.1)$$
and $$ds_2^2=a^2dt^2-a^2\cosh^2 t \ dx_1^2, \ n=3.  \eqno(7.2)$$
Using the geometric properties of our parameterization, it is possible to prove that the de Sitter spacetime has constant sectional curvature and it is possible to obtain the general relations $$R_{ij} + \dfrac{n-2}{a^2} g_{ij}=0. \eqno(7.3)$$ According to this result,   the curvature scalar $R$ and the cosmological constant $\Lambda$  fulfill the Einstein field equations in  absence of matter, that is: $$R_{ij}-\dfrac{1}{2}Rg_{ij}+\Lambda g_{ij}=0. \eqno(7.4)$$

They are $$ R=-(n-1)(n-2) \dfrac{1}{a^2}; \ \Lambda=-\dfrac{(n-2)(n-3)}{2}\dfrac{1}{a^2}. \eqno(7.5)$$
It is worth noticing  that  the nature of  cosmological constant can be  related to a centro-affine invariant derived from the  Tzitzeica surfaces.\\  
In our parametrization it is easy to check that the de Sitter spacetime is a Minkowski-Tzitzeica hypersurface whose invariant is $$  \dfrac{K^M_f \left( p\right)}{({d}_{f}^{M}\left( p\right))^{n+1}}=-\dfrac{1}{a^{2n}},  \eqno(7.6)$$
and then  the cosmological constant is
$$  \Lambda =-\dfrac{(n-2)(n-3)}{2} \sqrt[n]{\left|\dfrac{K^M_f \left( p\right)}{({d}_{f}^{M}\left( p\right))^{n+1}}\right|}. \eqno(7.7)$$
In other words, we have proven  that the nature of  cosmological constant is related to the property of  volume preservation  and it is not related to the constant sectional curvature, because it is known that there exist surfaces having constant Gaussian curvature which are not Tzitzeica surfaces.
In conclusions, cosmological constant can be fully recovered  from affine geometry arguments and this result could help in the debate of establishing the fundamental nature of this cosmic ingredient.

However, an important remark is necessary at this point. The approach reported here is completely classical. We did not tackle the problem of quantum origin of cosmological problem
\cite{Weinberg} and the possibility to explain the huge amount of energy related with it at the various epochs. Despite of this, the aim of this study is to point out that the "physical" meaning of cosmological constant can be recovered thanks to the fact that gravitational field emerges also  in Minkowski affine spaces without masses and in absence of standard matter. 
This feature can be extremely relevant in understanding the  nature of such cosmic fundamental ingredient.

\medskip

\section*{Acknowledgments}
S. C.   acknowledges the support of  INFN ({\it iniziative specifiche} MOONLIGHT2 and QGSKY) and the hospitality of the Department of Mathematics of Transilvania University of Brasov.
This paper is based upon work from COST action CA15117 (CANTATA), supported by COST (European 
Cooperation in Science and Technology).

\end{document}